\documentclass[11pt]{article}

\usepackage{amsthm}
\usepackage[english]{babel}
\usepackage[text={17cm,24cm},centering]{geometry}
\usepackage{graphicx}
\graphicspath{{./Figures/}}
\usepackage{epstopdf}
\usepackage[square,numbers]{natbib}
\usepackage{authblk}
\usepackage[format=hang,indention=-3em,textfont={small,it},labelfont={small,bf},width=0.92\textwidth]{caption}
\usepackage{enumitem}
\usepackage[notref,final]{showkeys}  

\usepackage{siunitx}
\newtheorem{remark}{Remark}

\usepackage{amssymb,MnSymbol,amsmath}
\usepackage{accents}
\usepackage{dsfont,mathbbol,upgreek}

\newcommand{\bI}{\boldsymbol{I}}

\newcommand{\bE}{\boldsymbol{E}}

\newcommand{\bLam}{\boldsymbol{\Lambda}}
\newcommand{\bal}{\boldsymbol{\alpha}}
\newcommand{\bC}{\boldsymbol{\mathcal{C}}}
\newcommand{\eee}{\: = \:}

\newcommand{\shh}{\!=\!}

\newcommand{\bT}{\boldsymbol{T}}

\newcommand{\bmT}{\boldsymbol{\mathcal{T}}}

\newcommand{\hh}{\hspace*{0.3pt}}

\newcommand{\nes}{\hspace*{-0.7pt}}

\newcommand{\opsi}{\overline{\psi}}

\DeclareMathOperator*{\argmin}{argmin}

\begin{document}

\title{A simple tool for the optimization of 1D phononic and photonic bandgap filters}

\author[1]{Prasanna Salasiya}
\author[1]{Bojan B. Guzina\thanks{Corresponding author (guzin001@umn.edu)}} 
\affil[1]{\small{Civil, Environmental, \& Geo- Engineering, University of Minnesota Twin Cities, Minneapolis, MN, U.S.}}
\maketitle
\begin{abstract}

\noindent We develop an effective computational tool for simulating the scattering of one-dimensional (1D) waves by a composite layer architected in an otherwise homogeneous medium. The layer is designed as the union of segments cut from various mother periodic media, which allows us to describe the wavefield in each segment in terms of the ``left'' and ``right'' (propagating or evanescent) Bloch waves. For a given periodic medium and frequency of oscillations, the latter are computed by solving the quadratic eigenvalue problem (QEP) which seeks the (real- or complex-valued) wavenumber -- and affiliated eigenstate -- of a Bloch wave. In this way the scattering problem is reduced to a low-dimensional algebraic problem, solved via the transfer matrix approach, that seeks the amplitudes of the featured Bloch waves (two per segment), amplitude of the reflected wave, and that of the transmitted wave. Such an approach inherently caters for an optimal filter (e.g. rainbow trap) design as it enables rapid exploration of the design space with respect to segment (i) permutations (with or without repetition), (ii) cut lengths, and (iii) cut offsets relative to the mother periodic media. Specifically, under (i)--(iii) the Bloch eigenstates remain invariant, so that only the transfer matrices need to be recomputed. The reduced order model is found to be in excellent agreement with numerical simulations. Example simulations demonstrate 40x computational speedup when optimizing a 1D filter for minimum transmission via a genetic algorithm (GA) approach that entails $O(10^6)$ trial configurations. Relative to the classical rainbow trap design where the unit cells of the mother periodic media are arranged in a ``linear'' fashion according to their dispersive characteristics, the GA-optimized (rearranged) configuration yields a $40\%$ reduction in filter transmissibility over the target frequency range, for the same filter thickness.
\end{abstract}

\maketitle

\section{Introduction}

\noindent Wave motion in periodic media has garnered significant research interest since early in the last century due to its relevance to various macroscopic phenomena such as wave dispersion, negative index of refraction, and band gaps~\citep{bensoussan2011, kushwaha1994, ziman2001}. In short, band gap refers to a frequency range where the incident waves are exponentially attenuated (more specifically reflected backward) due to destructive interference of the scattered wavefield; periodic media that are capable of realizing such exotic behavior in the sub-wavelength regime are commonly referred to as metamaterials~\citep{liu2011, muhammad2022}. Many applications of metamaterials such as negative refraction, superlens, cloaking, energy harvesting, vibration isolation and seismic protection, rely on the location and width of the band gaps~\citep{hussein2014, banerjee2019, goh2019}. Notwithstanding the fact that the wave manipulation devices designed for such purposes are necessarily bounded (i.e. compactly supported) and possibly aperiodic, as in the case of rainbow traps~\citep{shen2011a, shen2011b}, their performance is often interpreted through the lens of the frequency-wavenumber dispersion relationship computed for an \emph{infinite} periodic medium. This dichotomy inherently necessitates numerical or physical simulations of the full scattering problem, even for simple 1D systems, to achieve optimal design of metamaterial-based wave manipulation devices. As noted in~\cite{hussein2006}, having at least three to four unit cells is necessary to reflect -- at least to some degree -- the dispersive characteristics of an infinite periodic medium; however many metamaterial systems, such as rainbow traps, entail single unit cells extracted from various periodic media, rendering traditional methods of interpretation inadequate.

The scarcity of analytical tools for bounded, aperiodic metamaterial systems poses significant challenges for the optimal design of wave-manipulation devices. In the context of 1D problems, the existing literature predominantly focuses on either bounded cutouts of a single periodic medium, or very specific filter designs. For instance, the authors in~\cite{esquivel1994,bradley1994} deploy a transfer matrix method combined with the Bloch analysis to examine the reflectivity of a finite segment of a 1D periodic medium. Similarly, \cite{banerjee2017} focuses on the analysis of linearly-graded internal resonators in finite 1D chains, while \cite{hu2021} deals with metamaterial beams with \emph{a priori} chosen nonlinear gradation of local resonators via the spectral element method catering for flexural waves. However, these techniques are ill-suited for batch simulations targeting large sets of trial system configurations, highlighting the need for a universal reduced-order model that can be used as a computational backbone for the optimal design of more complex 1D metamaterial systems.

\begin{figure}[h!]
\centering{\includegraphics [width=1.0\textwidth ]{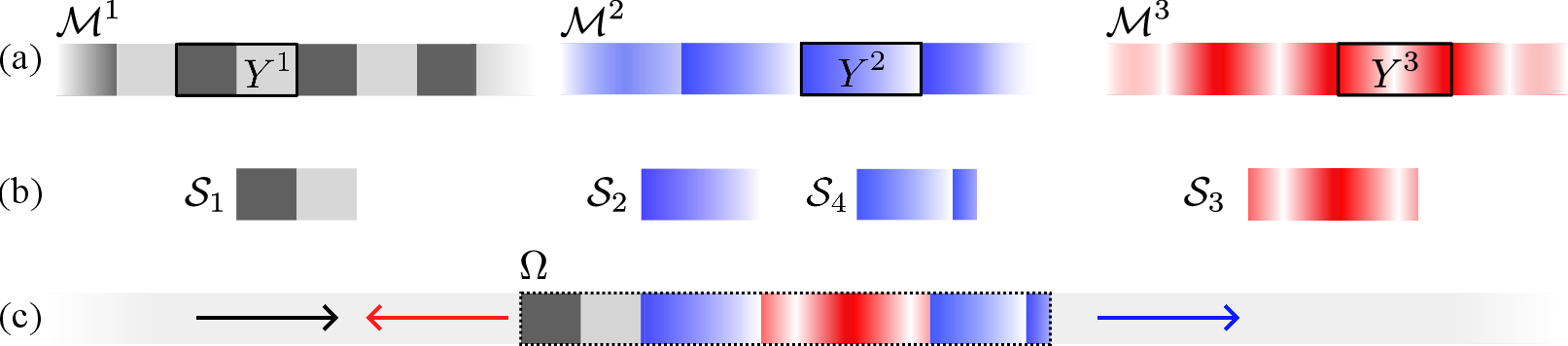}} 
\caption{(a) 1D mother periodic media $\mathcal{M}^q$ ($q\!=\!\overline{1,3}$) featuring the respective unit cells~$Y^q$; (b) segments $\mathcal{S}_j$ ($j\!=\!\overline{1,4}$), and (c) composite layer~$\Omega$ in otherwise homogeneous unbounded medium, architected as union of the bonded segments~$\mathcal{S}_j$.} \label{fig:segments}
\end{figure}

By building on the recent works in~\cite{Shahraki2022,salasiya2024}, this paper aims to address the gap by developing a semi-analytical tool to effectively compute the effects of 1D scattering by an aperiodic layer in an otherwise homogeneous infinite medium (see Fig.\ref{fig:segments}(c)), relevant to e.g. rainbow trapping or energy harvesting, using a combination of the Bloch wave analysis~\cite{Floq1883, Bloch1929, brillouin1946} and transfer matrix approach~\cite{lin1969, shen2000, cornaggia2020}. Specifically, our objective is to enable rapid exploration of the design space of a broad class of 1D (passive, metamaterial-based) wave manipulation devices. To this end, we assume that the architected layer is constructed as a 1D sequence of bonded \emph{strips} of dissimilar mother periodic media, see Fig.~\ref{fig:segments}. In our approach the wave motion in each mother periodic medium is precomputed, in terms of the ``left'' and ``right'' (propagating or evanescent) Bloch waves, by solving a quadratic eigenvalue problem (QEP)~\cite{tisseur2001} which seeks the wavenumber -- and affiliated eigenfunction -- of a Bloch wave for a given frequency of wave motion. While the linear eigenvalue problem (LEP), seeking the frequency of a Bloch wave for a given wavenumber, is typically used to compute the dispersion relationship~\cite{Lack2019}, the QEP offers a distinct  advantage for it enables computation of the wave motion \emph{inside} a band gap, that is essential for applications such as rainbow trapping and energy harvesting. In the context of the 1D wave equation, it was shown explicitly in~\cite{Shahraki2022} that the QEP yields exactly two (real- or complex-valued) wavenumbers for a given periodic medium. This allows us to express the solution of the wave equation with periodic coefficients, \emph{\'a la} d'Alembert, via superposition of the left- and right-going Bloch waves for an arbitrary segment of a periodic medium -- irrespective of its length and boundary conditions. With such result in place, 1D scattering by an architected layer is then computed via the transfer matrix approach used as a tool to ``stitch'' the segments of different mother periodic media. The key advantage of such an approach lies in the fact that the QEP (solved independently for each mother periodic medium) -- the most computationally taxing step -- can be \emph{precomputed} and used to rapidly solve the scattering problem for numerous combinations of the segment lengths, offsets, and permutations (with or without repetition). For each such trial architecture, the only item that needs recomputing are the transfer matrices which are then used to solve for the amplitudes of the germane Bloch waves (two per segment), amplitude of the reflected wave, and that of the transmitted wave. Thanks to its efficiency, this approach caters for the optimal design of 1D wave manipulation devices that inherently necessitates, irrespective of the minimization algorithm deployed, probing the design space via simulations performed over a significant number of trial system configurations. In this work, we deploy a genetic algorithm approach~\citep[e.g.][]{jian2022, liu2023} to illustrate the utility of the proposed approach toward an optimal design of a rainbow trap system. In the featured example, we achieve a 40x computational speedup and 40\% reduction in transmissibility of the optimized system relative to the conventional, i.e. ``linear'', rainbow trap design.

\section{Wave motion in infinite periodic medium}

\noindent Consider the time-harmonic wave motion at frequency~$\omega$ governed by the one-dimensional wave equation 
\begin{equation}\label{Scalarwave}
\frac{\text{d}}{\text{d} x} \left( G(x) \: \frac{\text{d} u}{\text{d} x} \right) + \omega^2 \rho(x) \: u \eee 0, \quad x \in \mathbb{R},
\end{equation}
where $G>0$ and $\rho>0$ are real-valued $Y$-periodic functions;
\[
Y = \{ x : 0 < x < \ell \}
\]
is the unit cell of periodicity, and the wavefield $u(x)$ carries implicit time dependence $e^{-i \omega t}$. In the context of phononics, $G$, $\rho$ and $u$ represent respectively the shear modulus, mass density, and transverse displacement. For photonics-relevant problems, $G$ is set to 1, while $\rho$ and $u$ denote the reciprocal of the squared speed of light in the material and electric field, respectively. 

On recalling the Floquet-Bloch theorem \citep{Floq1883, Bloch1929}, we seek the Bloch wave solution of~\eqref{Scalarwave} given by 
\begin{equation}\label{FB}
u(x) = \phi(x) \: e^{i k x}, \quad \phi : Y \mathrm{-periodic}
\end{equation}
where $k \in \mathbb{C}$ is the wavenumber, and $\phi$ implicitly depends on $\omega$ and $k$. By way of~\eqref{Scalarwave}--\eqref{FB}, we obtain the (periodic) boundary value problem supported on a unit cell, namely
\begin{align}\label{FBwave}
\begin{aligned}
\frac{	\text{d} }{\text{d} x_k} \left( G(x) \: \frac{	\text{d} \phi}{\text{d} x_k} \right) + \omega^2 \rho(x) \: \phi \eee 0, \quad x \in Y, \\
\phi|_{x=0} = \phi|_{x=\ell}, \qquad G \left. \frac{\text{d} \phi}{\text{d} x_k} \right| _{x=0} \!= G \left. \frac{\text{d} \phi}{\text{d} x_k} \right| _{x=\ell}, 
\end{aligned}
\end{align}
where $\text{d} / \text{d} x_k := \text{d} / \text{d} x + i k$.

\subsection{Quadratic eigenvalue problem (QEP)}

\noindent To facilitate the analysis, we introduce the Hilbert spaces  
\begin{eqnarray*}
    L_p^2(Y)  &\!\shh \!&  \{g\in L^2(Y):\, (g, g) < \infty, \,\,g|_{x=0} = g|_{x=\ell}\}, \\
    H^1_{p} (Y) &\! \shh \!& \left\{ g\in L^2(Y):\, \frac{\text{d} g}{\text{d} x} \in L^2(Y), \,\,g|_{x=0} = g|_{x=\ell} \right\},
\end{eqnarray*}
where $(g, h)=\int_Y g \; \overline{h} \; \text{d} x$ and $\overline{h}$ is complex conjugate of $h$. For given $\omega \in \mathbb{R}^+$,~\eqref{FBwave} can be rewritten in a weak form as the quadratic eigenvalue problem \citep{laude2009, Sriva2020}
\begin{equation}\label{QEP}
\int_Y \big( G \: \frac{\text{d} \phi}{\text{d} x} \: \frac{\text{d} \opsi}{\text{d} x} - \omega^2 \rho \: \phi \: \opsi \: \big) \text{d} x \,+\, k \int_Y i \: G \big( \phi \: \frac{\text{d} \opsi}{\text{d} x} - \opsi \: \frac{\text{d} \phi}{\text{d} x} \big) \text{d}x \,+\, k^2 \int_Y G \: \phi \: \opsi \: \text{d} x \eee 0, \quad \forall \: \psi \in H_p^1(Y)
\end{equation}
featuring the eigenvalues $k \in \mathbb{C}$ and eigenfunctions $\phi \in H_p^1(Y)$. In~\cite{Shahraki2022}, it was demonstrated that \eqref{QEP} has precisely two roots $k_{m}\!\in \mathbb{C}$, $m\!=\!\overline{1,2}\,$ within the first Brillouin zone. In situations where $k_{m}\!\in\mathbb{R}$, we arrange these two roots so that $k_{1}\!>0$ and $k_2\!<0$, which identifies the respective Bloch waves as the right- and left-propagating modes. For $k_{m}\!\notin\mathbb{R}$, on the other hand, we adopt the arrangement $\Im{(k_{1})}>0$ and $\Im{(k_{2})}<0$ which correspond respectively to the right- and left-decaying (evanescent) Bloch waves.

\begin{remark}
If in~\eqref{FBwave} we prescribe~$k\in\mathbb{R}$ in lieu of~$\omega\in\mathbb{R}^+$, we obtain the so-called linear eigenvalue problem (LEP) whose solution yields a countable infinity of eigenfrequencies, $\omega_n(k)\! \in \mathbb{R}^+$ ($n \!\in\! \mathbb{Z}^+$). The resulting dispersion diagram is known to be $\mathfrak{B}$-periodic  \cite{Shahraki2022}, where 
\[
\mathfrak{B} = \big\{k\in\mathbb{R}\!: k \in \big( -\pi, \pi \big] \big\}
\]
is the first Brillouin zone. In this vein, our QEP analysis implicitly assumes $\Re(k)\in\mathfrak{B}$. For given $n$ and $k \in \mathfrak{B}$, the open set $\omega_n(k)$ signifies the $n$th pass band $P_n\nes\subset\mathbb{R}^+$, while the set of contiguous open intervals $\mathbb{R}^+ \backslash \bigcup_{n \in \mathbb{Z}^+} \overline{P}_n $ gives the union of all band gaps.
\end{remark}


\subsection{The key result}\label{sec:QEP_prop}

\noindent The properties of the QEP eigenspectrum $\big\{ k_n(\omega), \phi_n (\cdot\; ; \omega) \big\}_{n=1}^{2}$ depend on the characteristic frequency regimes as described in~\cite{Shahraki2022} and are conveniently summarized below.

\begin{itemize}
\item When $\omega$ is in a pass band, $k_1 \!= \!-k_2 \in (0, \pi)$ which results in a pair of left- and right-\emph{propagating Bloch waves} as examined earlier. In this case the two eigenfunctions are complex conjugates of each other, i.e. $\phi_1 = \overline{\phi_2}$.
    
\item When $\omega$ is inside a band gap, $k_1 \!= \overline{k_2}$ with $\Re(k_1)=\Re(k_2) \in \{0,\pi\}$. This results in a pair of left- and right-\emph{evanescent Bloch waves}. Here, each eigenfunction is characterized by at least one node, namely a point within~$Y$ where $|\phi_{n}\!=0|$, $n\!=\!\overline{1,2}$.
    
\item If $\omega$ resides at the edge of the band gap (the so-called exceptional point~\cite{Heiss2012}), one obtains $k_1\! = k_2 \in \{ 0, \pi\}$ which results in a single \emph{standing Bloch wave}. More specifically, one encounters a degenerate situation where $\phi_1$ and $\phi_2$ each have at least one node and are linearly dependent.

\item If $\omega$ resides at the intersection between two dispersion branches (i.e.~Dirac point~\cite{ashraf2015}) -- which for 1D problems governed by~\eqref{Scalarwave} can occur only for $k_1\!= k_2\! \in\{0, \pi\}$ \cite{Shahraki2022}, the solution is given by a pair of  \emph{propagating Bloch waves} with linearly independent eigenfunctions $\phi_{n}$, $n\!=\!\overline{1,2}$.
\end{itemize}

In this setting, the key result from~\cite{Shahraki2022} demonstrates that for given $\omega\in\mathbb{R}^+$, a linear combination of the ``left''  and ``right'' Bloch waves 
\begin{equation} \label{dalbloch}
u(x) \:=\: A\, \phi_1(x) e^{i k_1 x} + B\, \phi_2(x) e^{i k_2 x}   
\end{equation}
with suitably chosen $A,B\in\mathbb{C}$ generates the \emph{complete solution}
of a boundary value problem governed by~\eqref{Scalarwave} over an \emph{arbitrary cutout} of the reference periodic medium and subject to \emph{arbitrary boundary conditions}. In what follows, we refer to~\eqref{dalbloch} as the d'Alembert-Bloch solution for it features Bloch waves, yet it resembles the classical d'Alembert solution of the 1D wave equation with constant coefficients for it decomposes the solution in to the ``left-going'' and ``right-going'' wavefields. 

\section{Scattering problem}


\noindent Motivated by the optimal design of 1D wave manipulation systems, we next consider a set of $Q$ mother periodic media $\mathcal{M}^{q}$ characterized respectively by the $Y^{q}$-periodic shear modulus~$G^{(q)}(\xi)$, mass density~$\rho^{(q)}(\xi)$, and unit cell 
\begin{equation}\label{unit cell}
Y^{q} = \big\{ \xi : 0\leqslant \xi \leqslant \ell^{(q)}\big\}, \qquad q=\overline{1,Q}
\end{equation}
whose cutouts are used to construct a 1D filter (see Fig.~\ref{fig:segments}). In the sequel, we shall denote the respective eigenspectra of~$\mathcal{M}^{q}$ by $\{k_n^{(q)},\phi_n^{(q)}\}_{n=1}^2$. 

\begin{figure}
\centering{\includegraphics [width=1.0\textwidth ]{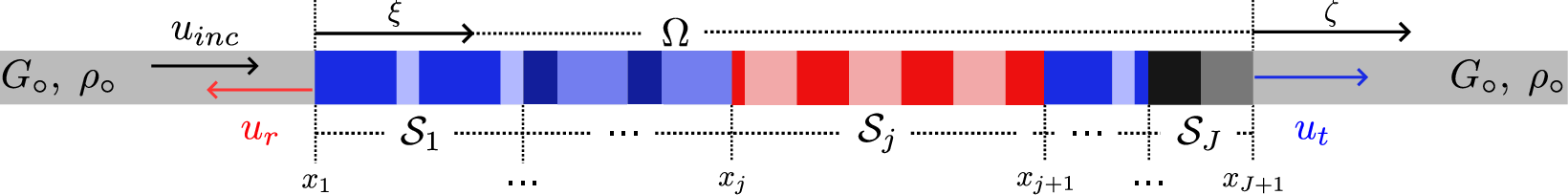}} 
\caption{1D wave scattering by an architected layer $\Omega$ sandwiched between two semi-infinite homogeneous domains. For simplicity, each segment within~$\Omega$ is rendered as being cut from a bilaminate mother periodic medium; this is not required by the analysis, see Fig.~\ref{fig:segments}(c) for a generic configuration.} \label{fig:trap}
\end{figure}

Letting $x_1=0$ and
\begin{equation}
    x_{j+1}=x_j+w_j, \qquad j=\overline{1,J} 
\end{equation}
for some $w_j\!>\!0$, we next consider $J$ dissimilar segments
\begin{equation}\label{segments}
 \mathcal{S}_j = \big\{ \xi : x_j \leqslant \xi \leqslant x_{j+1} \big\}, \qquad j=\overline{1,J} 
\end{equation}
that are $w_j$-wide and cut from the $s_j$-\emph{shifted} mother periodic media $\mathcal{M}^{{q_j}}$ (${q_j}\!\in\overline{1,Q}$)  in that 
\begin{equation} \label{matpro}
G_j(\xi) = G^{{(q_j)}}(\xi-(x_j+s_j)), \qquad  \rho_{j}(\xi) = \rho^{{(q_j})}(\xi-(x_j+s_j))  \qquad \textrm{for}~ \xi\in\mathcal{S}_j, 
\end{equation}
as shown in Fig.~\ref{fig:trap}. On denoting the eigenspectrum of the $j$th segment by $\{\kappa_n^{(j)},\varphi_n^{(j)}\}_{n=1}^2$, one finds that 
\[
\kappa_n^{(j)} = k_n^{(q_j)}, \quad \varphi^{(j)}_n(\xi) = \phi^{(q_j)}_n(\xi-s_j), 
\]
namely that the effect of shifting by $s_j$ the ``left'' cut of $\mathcal{M}^{{q_j}}$ leaves the eigenvalues unchanged, but induces translation of the respective ($Y^{q_j}$-periodic) eigenfunctions.

\begin{remark}
We allow different segments to be cutouts from the same periodic medium, in that ${q_j} \!= q_l$ for some $j\neq l$ and $j,l \in \overline{1,J}$. For generality, we also let $w_j \gtreqlessslant \ell^{({q_j})}$, whereby the segments need not be unit cell-wide.
\end{remark}


With the above definitions in place, we consider the scattering of of 1D scalar waves by an architected layer
\[
\Omega = \bigcup_{j=1}^J \mathcal{S}_j,
\]
embedded in an otherwise homogeneous medium endowed with shear modulus~$G_\circ$ and mass density~$\rho_\circ$. With reference to Fig.~\ref{fig:trap}, the system is subjected to an incident plane wave 
\begin{equation} \label{uinc}
u^{inc}=e^{i k_{\circ} \xi}, \qquad  k_{\circ} = \frac{\omega}{c_{\circ}} \in \mathbb{R}^+ 
\end{equation}
impinging ``from the left'', where  $c_{\circ}=\sqrt{G_{\circ}/\rho_{\circ}}$ is the phase velocity of the background medium. On denoting the segment boundaries by
\[
\Gamma_j=\{ \xi:\xi=x_j \}, \quad j=\overline{1,J\!+\!1}
\]
the continuity of displacements~$u$ and stresses $\sigma$ across $\Gamma_j$ require that 
\begin{equation}\label{cont}
u|_{\Gamma_j^-} = u|_{\Gamma_j^+} \quad\text{and} \quad\sigma|_{\Gamma_j^-} = \sigma_{\Gamma_j^+}, \quad j=\overline{1,J\!+\!1} \qquad\text{where}~~~ \sigma = G(\xi)\frac{\text{d}u}{\text{d}\xi}.
\end{equation}
The formulation of the scattering problem can now be completed by enforcing the radiation condition at infinity in that the solution for $\xi\!>\!x_{J+1}$ admits only the ``right-going'' waves, namely 
\[
u(\xi)  = \mathfrak{t} \, e^{i k_\circ (\xi-x_{J+1})}, \qquad \xi> x_{J+1}
\]
for some $\mathfrak{t}\in\mathbb{C}$ which signifies the amplitude of the transmitted wave.

\subsection{Wavefield in $\Omega$: d'Alembert-Bloch solution}\label{wavelayer}

\noindent On recalling~\eqref{dalbloch}, the displacement field $u_j$ and stress  field $\sigma_j$ in the $j$th segment~$\mathcal{S}_j$ can be written as 
\begin{align}\label{usigmalayer}
\begin{aligned}
u_j(\xi) &= \sum_{n=1}^{2}
\alpha^{(j)}_n \, \varphi^{(j)}_n(\xi) \; e^{i \kappa^{(j)}_n \xi} \\
\sigma_j(\xi) &= \sum_{n=1}^{2}
\alpha^{(j)}_n \, \psi^{(j)}_n(\xi) \; e^{i \kappa^{(j)}_n \xi}, \quad \psi^{(j)}_n = G_j(\xi) \left(\frac{\text{d}}{\text{d}\xi} + i \hspace{0.5pt}\kappa_n^{(j)}\right) \varphi^{(j)}_n
\end{aligned}
~, \qquad \xi \in \mathcal{S}_j, \qquad j=\overline{1,J}.
\end{align}
These expressions can be condensed using matrix-vector notation as
\begin{equation}\label{matveclayer}
    \begin{bmatrix} u_j \\ \sigma_j \end{bmatrix}(\xi) \,=\, \bLam_j(\xi) \; \bE_j(\xi) \; \bal_j, \quad j=\overline{1,J}
\end{equation} \label{LamE}
where  
\begin{equation}
\bal_j \,=\, \begin{bmatrix}
\alpha_1^{(j)} \\*[1mm] \alpha_2^{(j)} 
\end{bmatrix}, \qquad 
\bLam_j(\xi) \,=\, \begin{bmatrix}
\varphi^{(j)}_1(\xi) &\varphi^{(j)}_2(\xi) \\*[1mm]
\psi^{(j)}_1(\xi) & \psi^{(j)}_2(\xi)
\end{bmatrix} \qquad \text{and} \quad
\bE_j(\xi) \,=\, \begin{bmatrix}
e^{i \kappa^{(j)}_1 \xi} & 0 \\*[1mm]
0 & e^{i \kappa^{(j)}_2 \xi}
\end{bmatrix}.
\end{equation}
For future reference, we also introduce the short-hand notations
\begin{equation}
\bLam_{j,p} \,:=\, \bLam_j(x_p), \qquad \bE_{j,p} \,:=\, \bE_j (x_{p}), \qquad p\in\{j,j\!+\!1\} 
\end{equation}
and
\begin{equation}
\boldsymbol{\mathcal{E}}_{\!j} \,:=\,
\bE_{j,j}^{-1} \, \bE_{j,j+1} \,=\, \bE_{j,j+1} \, \bE_{j,j}^{-1} \,=\,
\begin{bmatrix}
e^{i \kappa_j^{(1)} w_j} & 0 \\*[1mm]
0 & e^{i kappa_j^{(2)} w_j}  
\end{bmatrix},
\end{equation}
$w_j\!>\!0$ being the width of the segment~$\mathcal{S}_j$. 

\subsubsection{Transfer matrix solution}

\noindent By way of~\eqref{matveclayer}, the Cauchy data on the left boundary of segment $\mathcal{S}_j$ is related to that on its right boundary as
\begin{equation}
\begin{aligned}
&\begin{bmatrix}
u_j \\ \sigma_j
\end{bmatrix} ( x_j ) \,=\, 
\bLam_{j,j} \; \bE_{j,j} \; \bal_j, \\
&\begin{bmatrix}
u_j \\ \sigma_j
\end{bmatrix} (x_{j+1}) \,=\, 
\bLam_{j,j+1} \; \bE_{j,j+1} \; \bal_j 
\end{aligned}  
\quad  \implies \quad 
\begin{bmatrix}
u_j \\ \sigma_j
\end{bmatrix}(x_{j+1}) \,=\, 
\bLam_{j, j+1} \; \boldsymbol{\mathcal{E}}_{\!j} \; ( \bLam_{j, j} )^{-1} \; 
\begin{bmatrix}
u_j \\ \sigma_j
\end{bmatrix}( x_j ), \quad j = \overline{1,J}.
\end{equation}
This allows us to introduce a local transfer matrix
\begin{equation*}
\bT_j \,:=\, \bLam_{j, j+1} \; \boldsymbol{\mathcal{E}}_{\!j} \; ( \bLam_{j, j} )^{-1}, \quad j=\overline{1, J}
\end{equation*}
such that
\begin{equation}\label{TM}
    \begin{bmatrix}
        u_j \\ \sigma_j
    \end{bmatrix}(x_{j+1}) \,=\, \bT_j \; \begin{bmatrix}
        u_j \\ \sigma_j
    \end{bmatrix}( x_j ), \quad j=\overline{1, J}.
\end{equation}
For future reference, we also define the compound transfer matrix
\[
\bmT_{\!\!j} \,:=\, \bT_j \; \bT_{j-1} \; \cdots \; \bT_{2} \; \bT_1, \quad j=\overline{1, J}.
\]

\subsection{Wavefield in the background medium}

\noindent On denoting by $u_r$ and $u_t$ the wavefields that are respectively reflected and transmitted by~$\Omega$ into the background medium, the displacement and stress fields in the left homogeneous space (index $j\!=\!0$) and the right homogeneous half-space (index $j\!=\!J\!+\!1$) can be written as 
\begin{align}
\begin{aligned} 
&u_0 \,=\, u_{inc}+u_r, &\qquad &\sigma_0 \,=\, \sigma_{inc} + \sigma_r, \\
&u_{J+1} \,=\, u_t, &\qquad &\sigma_{J+1}  \,=\, \sigma_t,\end{aligned}
\end{align}
where $u_{inc}$ is given by~\eqref{uinc} and $\sigma_{inc} \!=\! G_\circ\hh\text{d}u_{inc}/\text{d}\xi$. Since both half spaces are homogeneous, we can write $u_r(\xi)= \mathfrak{r} \; e^{-i k_{\circ} \xi}$ and $u_t(\xi) = \mathfrak{t} \; e^{i k_{\circ} \xi}$ where $\mathfrak{r}\in\mathbb{C}$ and $\mathfrak{t}\in\mathbb{C}$ are the unknown reflection and transmission coefficients. Letting $S_{\circ}:=i G_{\circ} k_{\circ}$, and $L:=|\Omega| = \sum_{j=1}^J w_j$, one can write
\begin{align}\label{usigmahomo}
\begin{aligned}
&u_0( \xi ) \,=\, e^{i k_{\circ} \xi} + \mathfrak{r} \; e^{-i k_{\circ} \xi}, &\quad &\sigma_0( \xi ) \,=\, S_{\circ} \; e^{i k_{\circ} \xi} - S_{\circ} \; \mathfrak{r} \; e^{-i k_{\circ} \xi}, \\
&u_{J+1}( \zeta ) \,=\, \mathfrak{t} \; e^{i k_{\circ} \zeta}, &\quad &\sigma_{J+1}( \zeta ) \,=\, S_{\circ} \; \mathfrak{t} \; e^{i k_{\circ} \zeta},
\end{aligned}
\end{align}
where the transmitted wavefield is conveniently described in terms of the \emph{shifted coordinate}, $\zeta=\xi-L$, with the origin at $\xi=x_{J+1}$. By following the analysis in Sec.~\ref{wavelayer}, \eqref{usigmahomo} can be written in the matrix-vector form as
\begin{equation}\label{matvechom}
\begin{bmatrix}
u_0 \\
\sigma_0
\end{bmatrix}(\xi) \,=\, \bLam_{\circ} \; \bE_{\circ}(\xi) \; 
\begin{bmatrix}
\mathfrak{r} \\
1
\end{bmatrix}, \qquad~~ 
\begin{bmatrix}
u_{J+1} \\
\sigma_{J+1}
\end{bmatrix}(\zeta) \,=\, \bLam_{\circ} \; \bE_{\circ}(\zeta) \;  \begin{bmatrix}
0 \\
\mathfrak{t}
\end{bmatrix},
\end{equation}
where
\begin{equation*}
\bLam_{\circ} = \begin{bmatrix}
1 & 1 \\
-S_{\circ} & S_{\circ}
\end{bmatrix} \quad \text{and} \quad
\bE_{\circ}(\xi) = \begin{bmatrix}
e^{-i k_{\circ} \xi} & 0 \\
0 & e^{i k_{\circ} \xi}
\end{bmatrix}.
\end{equation*}
We note that $\bE_{\circ}$ has the same structure as $\bE_j$, and that $\bE_{\circ}(0) = \bI$ with $\bI$ denoting the $2 \times 2$ identity matrix.

\section{Solution of the scattering problem}

\noindent For clarity, we summarize the key results from the previous section, namely 
\begin{align}\label{summ}
    \begin{aligned}
            \begin{bmatrix}
            u_0 \\ \sigma_0
        \end{bmatrix}(\xi=0) &\,=\, \bLam_{\circ} \; 
        \begin{bmatrix}
            \mathfrak{r} \\ 1
        \end{bmatrix},  \\
        \begin{bmatrix}
            u_j \\ \sigma_j
        \end{bmatrix}(\xi=x_{j+1}) &\,=\, \bT_j \; 
        \begin{bmatrix}
            u_j \\ \sigma_j
        \end{bmatrix}(\xi=x_{j}), \quad j\,=\,\overline{1, J}, \\
        \begin{bmatrix}
            u_{J+1} \\ \sigma_{J+1}
        \end{bmatrix}(\zeta=0) &\,=\, \bLam_{\circ} \begin{bmatrix}
            0 \\ \mathfrak{t}
        \end{bmatrix}.
    \end{aligned}
\end{align}
On the basis of~\eqref{summ}, the solution of the scattering problem is obtained by enforcing continuity of the Cauchy data \eqref{cont} across  segment interfaces, i.e.
\begin{align}
    \begin{aligned}
        \bLam_{\circ} \; 
        \begin{bmatrix}
            \mathfrak{r} \\ 1
        \end{bmatrix} &\,=\, 
        \begin{bmatrix}
            u_1 \\ \sigma_1
        \end{bmatrix}(0), \\
        \bT_{j} \; 
        \begin{bmatrix}
            u_{j} \\ \sigma_{j}
        \end{bmatrix}( x_{j} ) &\,=\, 
        \begin{bmatrix}
            u_{j+1} \\ \sigma_{j+1}
        \end{bmatrix}(x_{j+1}), \quad j\,=\,\overline{1,J\!-\!1},  \\
        \bT_{\!J} \; 
        \begin{bmatrix}
                u_{J} \\ \sigma_{J}
        \end{bmatrix}( x_{J} ) &\,=\, \bLam_{\circ} \; 
        \begin{bmatrix}
            0 \\ \mathfrak{t}
        \end{bmatrix};
    \end{aligned}
\end{align}
a result can be condensed as 
\begin{equation}\label{TMsol}
    \bmT_{\!\!J} \; \bLam_{\circ} \; 
    \begin{bmatrix}
        \mathfrak{r} \\ 1
    \end{bmatrix} \,=\, \bLam_{\circ} \; 
    \begin{bmatrix}
        0 \\ \mathfrak{t}
    \end{bmatrix}.
\end{equation}
Once $\mathfrak{r}$ and $\mathfrak{t}$ are computed by solving the $2\times 2$ algebraic system~\eqref{TMsol}, the wavefield inside $\Omega$ -- as controlled by $\bal_j$ ($j\!=\!\overline{1,J}$) -- is readily obtained from the previous equations.

\subsection{Design parameters}

\noindent Let the architected layer $\Omega$ comprise $J$ segments that are cut from $Q \leqslant J$ mother periodic media $\mathcal{M}^q$, $q\!=\!\overline{1,Q}$. Then the design space that can be rapidly searched by the foregoing analysis includes 
\begin{itemize}
    \item the number of segments, $J$;
    \item segment makeups ${q_j}\!\in\overline{1,Q}\:$ ($j\!=\!\overline{1,J}$);
    \item segment widths $|\mathcal{S}_j|=w_j$ ($j\!=\!\overline{1,J}$);
    \item translations~$s_j$ of the mother periodic medium $\mathcal{M}^{q_j}$ relative to segment $\mathcal{S}_j$.
\end{itemize}
For given~$J$, the above design space is conveniently described by the configuration vector
\begin{equation} \label{convec}
\bC_{\Omega} :\,=\, \{ ({q_j}, \, w_j, \, s_j) \}_{j=1}^J.
\end{equation}

In the next section, our goal is to design $\Omega$ with the particular objective of minimizing transmission over a given frequency band, as in the case of rainbow trapping~\citep{shen2011a, shen2011b}. From the foregoing analysis, one finds that having $Q$ solutions of the QEP~\eqref{QEP} -- computed respectively for $\mathcal{M}^{q}$($q\!=\!\overline{1,Q}$) -- provide the basis for a rapid exploration of~$\bC_{\Omega}$. For example the handle over ${q_j}$, $j\!=\!\overline{1,J}$ corresponds to permutations (with or without repetition) of the $J$ segments. Similarly, $s_j$ allow us to cut out any given segment from its mother periodic medium starting from various ``micro''-locations within the unit cell. This parameter is important for it controls the local impedance contrast, and so the reflection and transmission, at  interfaces between the neighboring segments. Further if the excitation frequency resides at the edge of a band gap (namely the exceptional point), one could potentially engineer \emph{global} resonance of a system by simply choosing the \emph{local} parameter $s_j$ so it corresponds to the common node of the two eigenfunctions, $\phi_n(x)$ ($n\!=\!\overline{1,2}$), at the exceptional point \cite{Shahraki2022}. 


\section{Numerical results}

\noindent In what follows, reference numerical simulations are performed via NGSolve – an open-source, Python-based
finite element (FE) computational platform \cite{NGSolve}. For clarity, all physical parameters are rendered dimensionless via the characteristic length $L_{\circ}$ of the problem, the mass density of the background medium, and the shear modulus thereof. In what follows, all mother periodic media are taken as bilaminate in that 
\[
Y^{q}=\{\xi\!: 0<\xi\!<\ell^{(q)}\}, \qquad 
  (G^{(q)}(\xi),\rho^{(q)}(\xi)) \:=\: \left\{
\begin{array}{cc}
(G_1^{(q)},\rho_1^{(q)})  &  0 \, \leqslant \, \xi \,< \, \ell^{(q)}-h^{(q)},  \\*[1mm]
(G_2^{(q)},\rho_2^{(q)})  & \ell^{(q)}-h^{(q)} \, \leqslant \, \xi \, \leqslant \, \ell^{(q)}.
  \end{array}\right.
\]

\begin{table}
\caption{Material properties of the bilaminate periodic media featured by numerical examples. First row: QEP example, Sec.~\ref{sec:QEP}. Second row: rainbow trap (RT) example, Sec.~\ref{sec:ver}. Here, subscript 1 (resp.~2) refers to the properties of the first (resp. second) layer within the unit cell.} 
\vspace*{2mm}
\centering
\begin{tabular}{|c|c|c|c|c||c|c|} \hline
{Parameter}  & $G^{(q)}_1$ & $\rho^{(q)}_1$ & $G^{(q)}_2$ & $\rho^{(q)}_2$ & $\ell^{(q)}$ & $h^{(q)}$ \\ \hline\hline
\rule{0pt}{10pt} $q=1$, QEP example  &  1  &  1  &  0.1  &  1.6 & 1 & 0.5  \\ \hline
\rule{0pt}{10pt} $q=\overline{1,21}$, RT example &  0.5  & 4 & 0.5 & 2.5 & $\num{1.212e-2} + q\hh(\num{8.7e-4})$ & $\num{2.597e-2}$ \\ \hline
\end{tabular}
\label{table:ucprop}
\end{table}

\subsection{Quadratic eigenspectrum}\label{sec:QEP}
\begin{figure}[t]\vspace*{3mm}
\centering{\includegraphics [width=0.67\textwidth ]{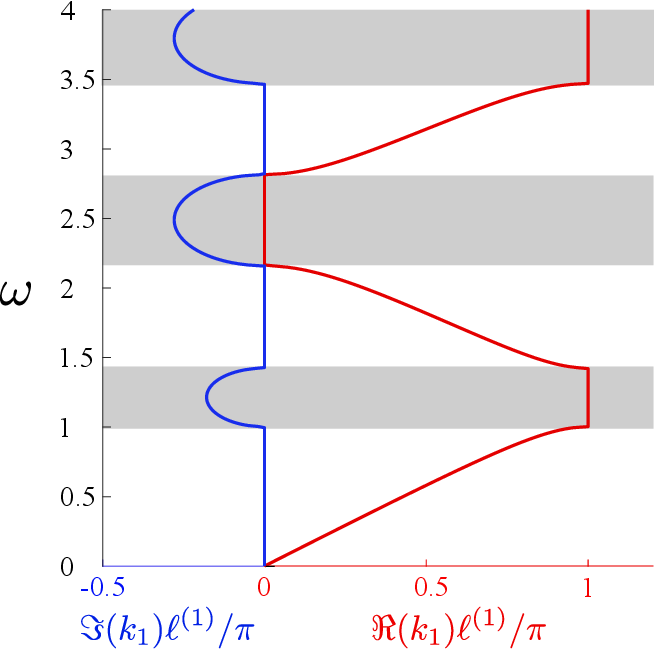}} 
\caption{Dispersion diagram of the bilaminate periodic medium whose properties are listed in Table~\ref{table:ucprop} (row 1). The real and imaginary parts of the wavenumber $k_1$, obtained by solving QEP~\eqref{QEP} for $\omega \in [0,3.8)$, are shown respectively in red and blue.} \label{fig:QEP}
\end{figure}

\noindent We start the illustrations with computation of the eigenspectrum $\big\{ k^{(1)}_n(\omega), \phi^{(1)}_n(\cdot\:; \omega) \big\}_{n=1}^2$ of QEP~\eqref{QEP} for the bilaminate periodic medium with properties listed in Table~\ref{table:ucprop} (row 1) by letting $L_\circ=|Y^{1}|$, $\omega \in [0,3.8)$, and meshing the unit cell with forty 1D elements of order $p=3$.

The eigenvalues are obtained by first converting \eqref{QEP} to a larger size linear eigenvalue problem \cite{Lack2019}, and then solving the latter via Arnoldi algorithm \cite{Leho1998} available through the NGSolve library. It was observed that 100 iterations of the Arnoldi solver are sufficient to obtain highly accurate eigenspectrum for the first ten dispersion branches. On recalling the fact that $k_n(-\omega)=k_n(\omega)$ (see Sec.~\ref{sec:QEP_prop}), the output of the QEP computations is shown in Fig.~\ref{fig:QEP} which plots the first eigenvalue $k_1\!\in\mathbb{C}$ over the positive half of the first Brillouin zone. Here it is useful to recall that within the band gap, $\Re(k_1)=\Re(k_2) \in \{0, \pi\}$ and $k_1=\overline{k_2}$ as examined in Sec.~\ref{sec:QEP_prop}.

\subsection{Verification of the transfer matrix solution}\label{sec:ver}
\begin{figure}
\centering{\includegraphics [width=0.97\textwidth ]{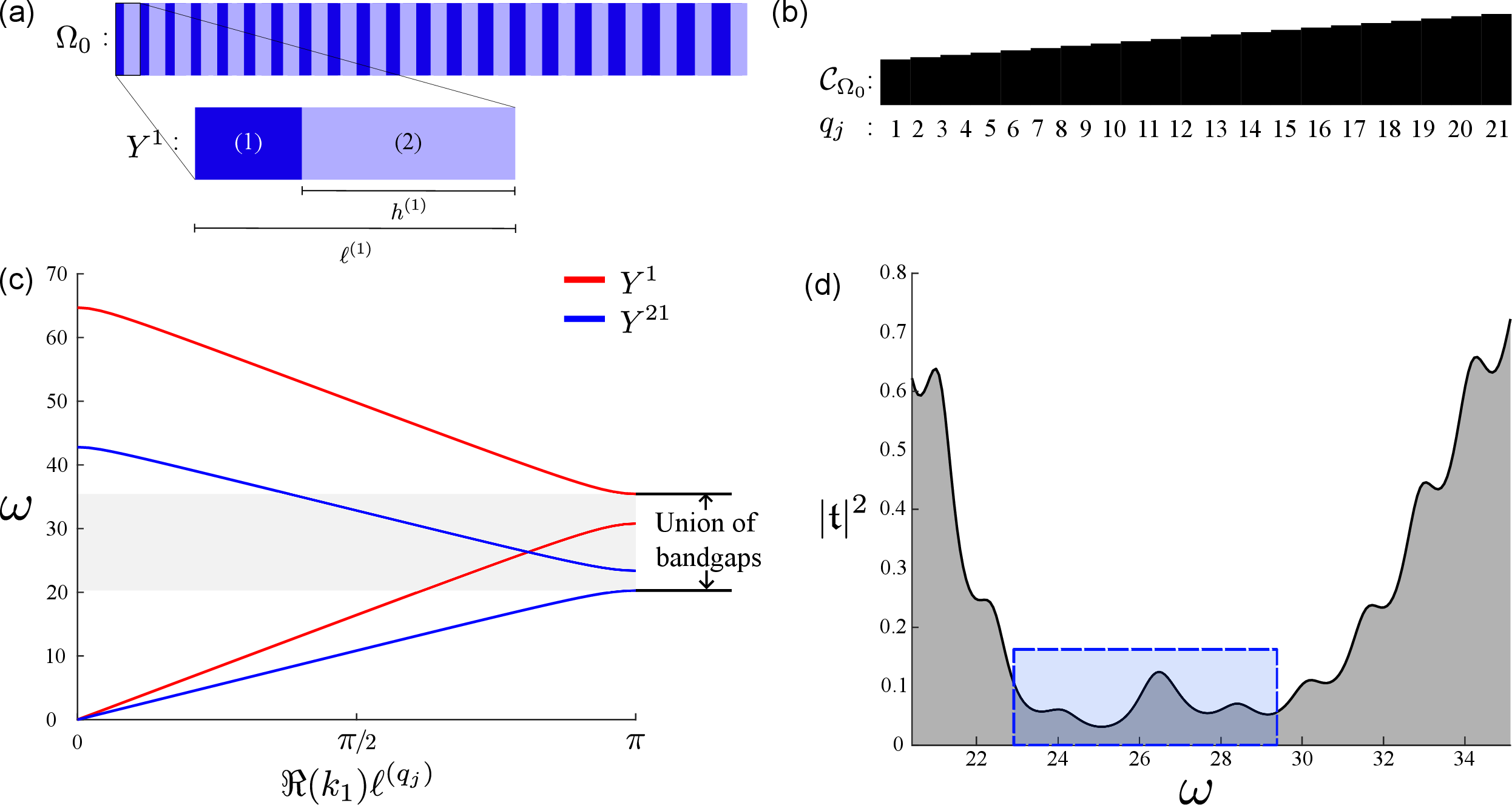}} 
\caption{(a) Initial rainbow trap design $\Omega_0$, unit length-wide and designed as the union of $21$ (unit cell-wide) segments of bilaminate mother periodic media $\mathcal{M}^q$ $(q=\overline{1,21})$; (b) schematic illustrating the configuration vector $\bC_{\Omega_0}$, with the height of each bar scaled to represent the segment makeup ${q_j}$; (c) the first two dispersion branches of the mother periodic media~$\mathcal{M}^{1}$ and~$\mathcal{M}^{21}$, with the union of the band gaps of $\mathcal{M}^q$ $(q=\overline{1,21})$ highlighted in gray; and (d) transmission-frequency diagram $|\mathfrak{t}|^2(\omega)$ due to $\bC_{\Omega_0}$, plotted over the union of band gaps $\omega \in [20.42, 35.12]$. The effective band gap created by $\Omega_0$, $\omega \in [22.9, 29.40]$, is highlighted in blue.} \label{fig:RT_disp}
\end{figure}

\noindent To verify the transfer matrix (TM) solution against full-field FE simulations, we construct a rainbow trap (Fig.~\ref{fig:RT_disp}(a)) with $J\!=\!21$ unit cell-wide segments~$\mathcal{S}_j$ extracted from the mother periodic media $\mathcal{M}^q$ ($q\!=\!\overline{1,21}$) such that $q_j\!=\!j$. The rainbow trap is designed following the usual approach \citep{shen2011a, shen2011b} where the target band gap of $\mathcal{M}^q$ is slightly shifted downward relative to that of $\mathcal{M}^{q-1}$. This is achieved by designing a chirped structure where the length of the second layer of a bilaminate unit cell is kept fixed, while the length of the first layer is increased linearly across the rainbow trap, see Table~\ref{table:ucprop} (bottom row) and Fig.~\ref{fig:RT_disp}(a). For this example, we take the length of the rainbow trap as the characteristic length of the problem $L_\circ$. With reference to~\eqref{convec}, Fig.~\ref{fig:RT_disp}(b) shows the schematic of the initial configuration vector, $\bC_{\Omega_0} \!=\! \{ (q_j\!=\!j, \, w_j\!=\!\ell^{(j)}, \, s_j\!=\!0)\}_{j=1}^{21}$, while Fig.~\ref{fig:RT_disp}(c) shows the resulting union of the (first) band gaps. Fig.~\ref{fig:RT_disp}(d) illustrates the performance of the rainbow trap as computed by the proposed approach, by plotting the squared amplitude of the transmission coefficient $|\mathfrak{t}|^2$ versus $\omega$, whose range in the panel corresponds to the union of the band gaps. As can be seen from the display, the effective band gap achieved by the filter is significantly more narrow than the union of the band gaps featured in Fig.~\ref{fig:RT_disp}(c).

\begin{figure}
\centering{\includegraphics [width=0.9\textwidth ]{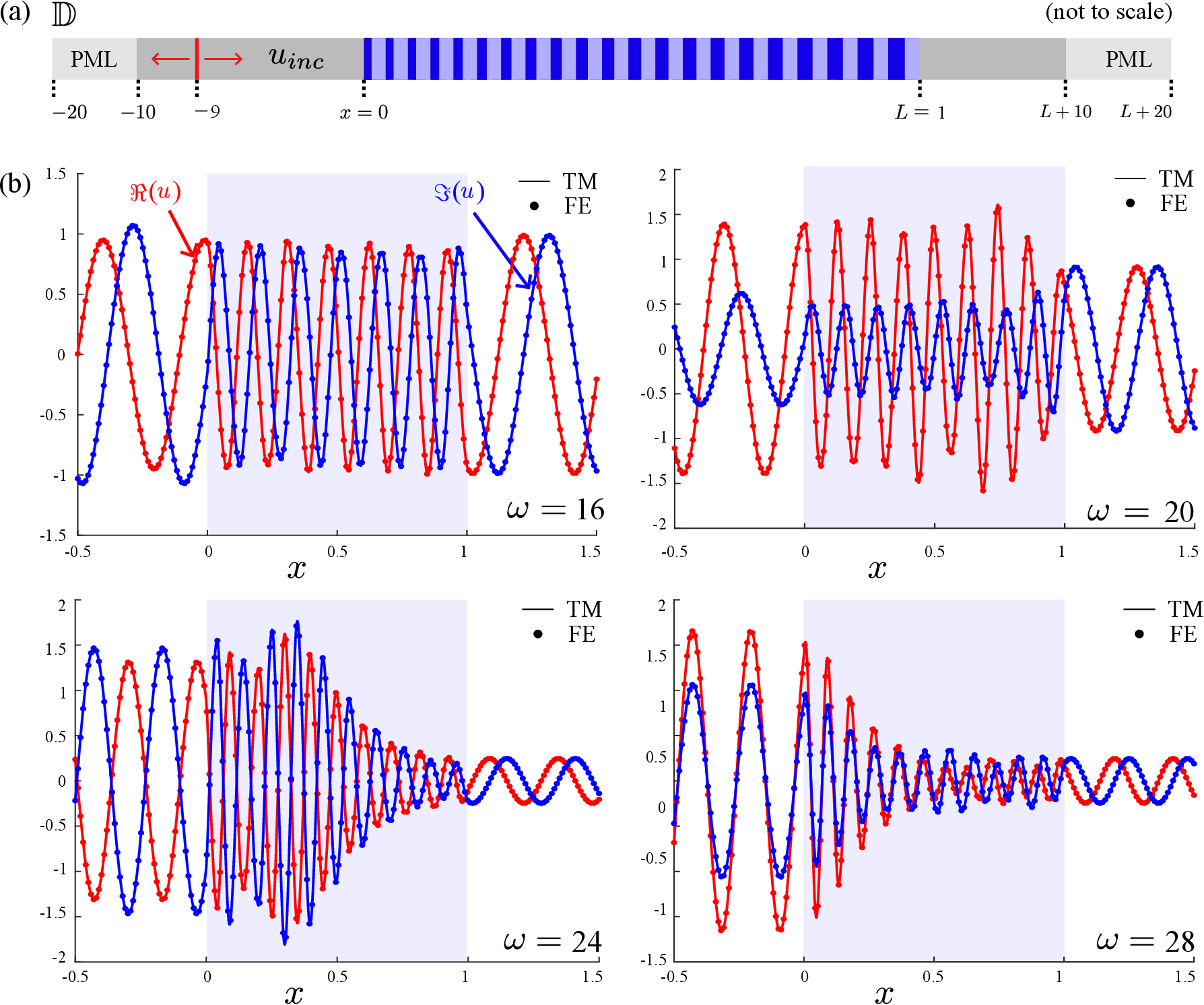}} 
\caption{(a) Simulation domain $\mathbb{D}$ containing the rainbow trap~$\Omega$ that is characterized by the configuration vector $\bC_{\Omega_0}$ and sandwiched between two homogeneous layers supported by a PML-type absorbing boundary. The red line at $x=-9$ indicates the location of the point source, $\delta(x+9)$, used to generate the incident plane wave. (b) comparison between the transfer matrix (TM) result and finite element (FE) simulations for $\omega \in\{16,20,24,28\}$.} \label{fig:RT_ver}
\end{figure}

To provide a reference numerical solution, time-harmonic FE simulations are performed in NGSolve to solve
\[
\frac{\text{d}}{\text{d} x} \left( G(x) \: \frac{\text{d} u}{\text{d} x} \right) + \omega^2 \rho(x) \: u \:=\: \delta(x-x_0), \quad x \in \mathbb{D},
\]
where $\mathbb{D}\supset \Omega$ is a truncated computational domain; Dirac delta function $\delta(x-x_0)$ is a point source at $x_0$ located ``to the left'' of the rainbow trap, used to generate the 1D incident plane wave; and $\omega$ is the excitation frequency. The computational domain $\mathbb{D}$ consists of the rainbow trap sandwiched between two homogeneous media ($G_\circ=1, \, \rho_\circ=1$), each followed by an absorbing boundary implemented via perfectly matched layers (PML) of length 10, see Fig.~\ref{fig:RT_ver}(a). The point source at $x_0=-9$ is approximated by Gaussian distribution with suitable variance $\gamma^2$. The simulation domain is meshed approximately with 1800 1D elements of order $p = 3$.

Fig.~\ref{fig:RT_ver}(b) compares the FE simulations versus the proposed transfer matrix (TM) approach at four excitation frequencies, $\omega \in \{ 16, 20, 24, 28 \}$. As can be seen from the display, the two sets of results are practically indistinguishable. The frequencies are selected such that $\omega=16$ is in the pass band of all mother periodic media, $\omega=20$ is at the edge of the band gap of mother medium~$\mathcal{M}^{21}$, whereas $\omega\in\{24,28\}$ reside within the union of the band gaps. The effect of rainbow trapping can be seen in Fig.~\ref{fig:RT_ver}(b), where the amplitude of the transmitted wave is severely diminished for $\omega\in\{24,28\}$. 

\subsection{Optimization of the rainbow trap}

\noindent The conventional design of rainbow trap systems operates under the assumption that the effective band gap achieved by a bounded filter $\Omega$ corresponds approximately to the union of the band gaps of the mother periodic media, $\mathcal{M}^j, \; j = \overline{1, J}$ (see Sec.\ref{sec:ver} and Fig.\ref{fig:RT_disp}(c)). However, for the filter shown in Fig.~\ref{fig:RT_disp}(a), as well as ``linear'' 1D rainbow trap designs available elsewhere in the literature~\citep{shen2011a, shen2011b}, there is a significant energy leakage close to the edges of the union of the component band gaps, see Fig.~\ref{fig:RT_disp}(d). This observation suggests that there is a potential for optimizing the filter, relative to its rainbow-trap ``linear'' design, to minimize the energy transmissibility 
\begin{equation} \label{transen}
\mathcal{T}(\bC_\Omega; \omega_1, \omega_2) := \int_{\omega_1}^{\omega_2} \big| \mathfrak{t}(\omega; \bC_\Omega) \big|^2 \; \text{d}\omega.
\end{equation}

\begin{figure}
\centering{\includegraphics [width=0.84\textwidth ]{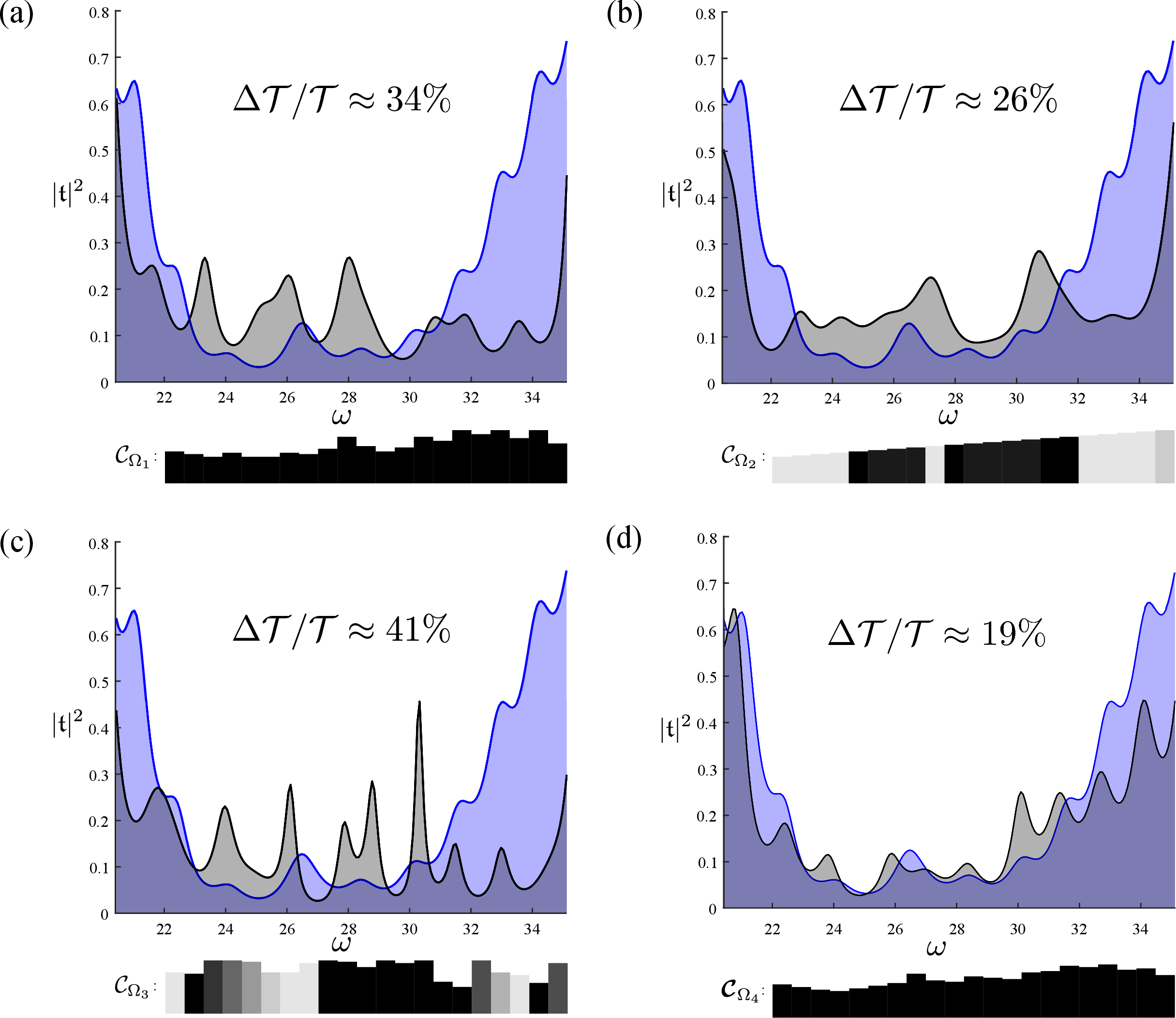}} 
\caption{Transmission diagrams achieved by the optimized (black) and initial (blue) filters, when the design space includes: (a) segment permutations with repetition; (b) window translations of individual segments; (c) both segment permutations (with repetition) and window translations, and (d) segment permutations without repetition. The inset at the bottom of each panel shows the schematic of the optimized configuration, with the bar heights representing segment makeups ${q_j}$ and the bar transparency indicating window translation $s_j$ ($100(p\!-\!1)\%$ bar transparency for $s_j\!=\!(p/10)w_j$).} \label{fig:RT_trans_comp}
\end{figure}

\paragraph{Segment permutations}
We begin the optimization of the unit-length ``linear'' configuration $\bC_{\Omega_0}$ shown in Fig.~\ref{fig:RT_disp}(a)--(b), using~\eqref{transen} as the cost function, by allowing for segment permutations. In this setup, a plane wave with unit amplitude is incident upon the left boundary of the system, and the corresponding transmission amplitude, $\mathfrak{t}$, is observed at the right boundary $\xi=x_{J+1}$. In this case, the design space spans all possible permutations \emph{with repetition} of the unit cells, yielding a total of $21^{21}$ individual configurations. With reference to~\eqref{unit cell}--\eqref{matpro}, we allow for the possibility that $q_j \!\in\! \{\overline{1, 21}\}$ for $j\!=\!\overline{1, 21}$, while keeping the segment widths unit cell-wide ($w_j\!=\nes\ell^{(q_j)}$) fixed and not permitting window translations ($s_j=0$). This ensures that the optimization focuses purely on the arrangement of the unit cells rather than any geometric alterations. With reference to Table~\ref{table:ucprop}, the optimization problem can thus be written as
\[
\bC_{\Omega_1} = \argmin_{q_j\in\{\overline{1, 21}\},\, j=\overline{1,21}} \mathcal{T} ( \bC_\Omega; \omega_1,\omega_2), \qquad \bC_\Omega = \big\{ (q_j,\, w_j\!=\nes\ell^{(q_j)}\nes,\, s_j\!=\!0) \big\}_{j=1}^{21} 
\]
where $\omega_1\!=\!20.42$ and  $\omega_1\!=\!35.12$, see Fig.~\ref{fig:RT_disp}. 
To effectively explore the extensive design space, we employ the genetic algorithm (GA) function available via Matlab Optimization Toolbox~\cite{Optimization}, utilizing the \texttt{ga} function with the \texttt{UseParallel} option enabled for parallel processing. The following GA options are found to be suitable for our computational needs:
\begin{itemize}
\item \texttt{Populationsize} : 5000
\item \texttt{EliteCount} : 100
\item \texttt{CrossoverFraction} : 0.8
\item \texttt{MaxStallGenerations} : 100
\item \texttt{FitnessScalingFcn} : \texttt{@fitscalingprop}
\item \texttt{MutationFcn} : \texttt{\{@mutationgaussian, 1, 0\}}
\end{itemize}
In what follows, the simulations are performed on a 16-core AMD 7763 processor with 32 GB of memory. To enhance the robustness of the optimization process, we execute 100 GA searches in parallel. Figure~\ref{fig:RT_trans_comp}(a) shows a comparison between the transmission curves achieved by the optimized and initial (i.e.~``linear'') rainbow trap design, featuring a relative reduction in the energy  transmissibility $\Delta\mathcal{T}/\mathcal{T}=34\%$. Here it is interesting to note that the length of the optimized filter $\bC_{\Omega_1}$ is 0.97, i.e. slightly less than that of the initial filter $\bC_{\Omega_0}$, with the length variability arising from the facts that: (i) we allow for permutations with repetition, and (ii) the reference unit cells have slightly different widths by design. 

\paragraph{Window translations}
We next update the design space to include only window translations $s_j \!\in \mathfrak{S}_j$ ($j=\overline{1,21}$) where
\[
\mathfrak{S}_j :=\; \big\{ \mathfrak{s}^{(p)}_j \!\!:\; \mathfrak{s}^{(p)}_j\!=\tfrac{p}{10}\, w_j, \quad p = \overline{0, 9} \big\},
\]
while maintaining the original makeups and widths of the segments~$\mathcal{S}_j$. In this setting, the minimization problem starts from initial configuration $\bC_{\Omega_0}$ and searches for $\bC_{\Omega_2}$ such that
\[
\bC_{\Omega_2}=\argmin_{s_j \in \mathfrak{S}_j \, j=\overline{1,21}} \mathcal{T} ( \bC_\Omega; \omega_1,\omega_2),  \qquad \bC_\Omega=\big\{ (j,\, w_j\!=\nes\ell^{(j)}\nes,\, s_j) \big\}_{j=1}^{21}.
\]
Using the same GA approach to explore the new design space, Fig.~\ref{fig:RT_trans_comp}(b) shows a relative reduction in the energy transmissibility $\Delta\mathcal{T}/\mathcal{T}=26\%$, achieved by~$\bC_{\Omega_2}$ with no change in the (unit) length of the filter. 

\paragraph{Segment permutations with window translations}
Lastly, we enlarge the design space by allowing for both segment permutations (with repetition) and window translations while maintaining the segments unit-cell-wide, namely 
\[
\bC_{\Omega_3}=\argmin_{q_j\in \{\overline{1,21}\},\, s_j \in \mathfrak{S}_j, \, j=\overline{1,21}} \mathcal{T} ( \bC_\Omega; \omega_1,\omega_2), \qquad \bC_\Omega=\big\{ (q_j, \, w_j\!=\nes\ell^{(q_j)}\nes,\, s_j) \big\}_{j=1}^{21}.
\]
As in previous minimization efforts, we explore the featured design space via genetic algorithm by executing 100 GA searches in parallel. In this case the optimized filter $\bC_{\Omega_3}$ carries an increased length of $1.2$ and achieves a $41\%$ reduction in the energy transmissibility, with the respective transmission diagrams being shown in 
Fig.~\ref{fig:RT_trans_comp}(c). As can be seen from the schematic of $\bC_{\Omega_3}$ shown at the bottom inset, the energy transmissibility is minimized by having the segment makeups dominated by the periodic media $\mathcal{M}^j$ featuring ``lower'' band gaps, i.e. those with higher index $j$. This feature is in fact responsible for the increase in filter length due to the fact that $|Y^{j+1}|\!>\!|Y^j|$, $j\!=\!\overline{1,20}$.


\subsection{Discussion}

\noindent With the use of a suitable cost function and search algorithm, the proposed computational framework can be easily adapted to include other constraints, for example (i) permutations without repetition, (ii) variable segment widths $w_j$ with an additional penalty on the overall filter thickness $\sum_{j=1}^{J} w_j$, and (iii) non-contiguous target band gaps, e.g. transmission suppression over $(\omega_1,\omega_2)\cup(\omega_3,\omega_4)$ for some $\omega_4>\omega_3>\omega_2>\omega_1$. For (i), we specifically have 
\[
\bC_{\Omega_4} = \argmin_{q_j\in\{\overline{1, 21}\},\, j=\overline{1,21}} \mathcal{T} ( \bC_\Omega; \omega_1,\omega_2), \qquad \bC_\Omega = \big\{ (q_j,\, w_j\!=\nes\ell^{(q_j)}\nes,\, s_j\!=\!0) \big\}_{j=1}^{21}, \quad \text{with } q_j \neq q_k \; \text{for} \; j \neq k. 
\]
In this case the GA can be suitably customized via the \texttt{\{`PopulationType', `custom'\}} option, available through the Matlab $\texttt{ga}$ function, to enforce the non-duplicity makeup constraint in the following way. 
\begin{enumerate}
\item \texttt{randperm(21)} is used to generate initial population pool of 5000 genes via random permutations of $ q_j, \; j=\overline{1,21}$.
\item To implement the crossover between two parents while ensuring that the offspring satisfies the non-duplicity constraint, a random crossover point is selected in the first parent. The child is then constructed by combining (i) the gene sequence from the first parent up to the chosen crossover, and (ii) the set difference between the second parent and the segment of the first parent preceding the crossover. The remaining parameters are left unchanged, i.e. \texttt{CrossoverFraction} is 0.8, \texttt{MaxStallGenerations} is 100, and \texttt{EliteCount} is 100.
\item Two randomly selected sites in the parent gene are swapped to generate a mutated child. This preserves non-duplicity of sites, i.e. makeups, within the child gene.
\end{enumerate}
Fig.~\ref{fig:RT_trans_comp}(d) shows the result of the above GA optimization, resulting in a 19\% decrease in energy transmissibility. As can be seen from the comparison with Fig.~\ref{fig:RT_trans_comp}(a), allowing for the repetition of certain mother periodic media (which leaves others out) in the design process leads to a significant improvement in the performance of the filter. 

In the above examples, optimized configurations $\bC_{\Omega_m}$ ($m\!=\!\overline{1,4}$) uniformly feature a reduction in the energy transmissibility~\eqref{transen} relative to~$\bC_{\Omega_0}$, albeit at a cost of somewhat increased energy transmission at interior frequencies removed from the edges of the target band gap. In situations where this ``inner'' frequency band is deemed essential for the performance of a filter, the integrand in~\eqref{transen} can be endowed with a suitable weight function~$\chi(\omega)$ that reflects such additional constraint. More generally, $\chi(\omega)$ can be designed to suppress energy transmission over non-contiguous frequency bands that was stipulated earlier.  

To quantify the computational advantages of the Bloch eigenstate-based transfer matrix (TM) approach, it is important to compare its performance with traditional FE simulations. On the computer used for TM calculations, a single-frequency FE simulation for a single filter configuration $\bC_{\Omega}$ takes approximately 3.3 seconds. By contrast, the upfront computational effort of solving the 3150 QEPs for  150 input frequencies (covering the target band gap) and 21 mother periodic media takes 26 minutes, upon which the TM method completes the simulation of wave transmission for a single  configuration $\bC_{\Omega}$ at 150 frequencies ($\omega \in [20.42, 35.12]$) within just 12 seconds. Letting $J\!=\!21$, for a set of $N_c$ configurations $\bC_{\Omega_m}$ comprised of the~$Q$ mother periodic media, the respective runtimes for $N_\omega$ test frequencies can accordingly be written as 
\begin{equation*}
\begin{aligned}
& \text{TM runtime:} && \text{$R_{\text{\tiny TM}}$[sec]} \;=\; N_\omega \big(0.495\,Q + 0.08\, N_c \big),  \\    
& \text{FE runtime:} && \text{$R_{\text{\tiny FE}}$[sec]} \;=\; 3.3\, N_\omega N_c.      
\end{aligned} 
\end{equation*}
This results in a significant speedup for a large number of filter configurations. Specifically, when $N_c$ is large the relative upfront cost of the TM simulations becomes negligible, and one obtains the speedup ratio
\[
\lim_{N_c\to\infty}\frac{R_{\text{\tiny FE}}}{R_{\text{\tiny TM}}} = \frac{3.3}{0.08} \;\approx\; 40. 
\]

In the above examples where the GA-based exploration of the design space entails up to 3 million trial simulations, this is indeed the speedup factor observed. In general, the latter is expected to grow significantly with either increasing filter length, or an ``upward'' shift in the frequencies covered by the target band gap. Each of these situations would require a larger FE mesh size and so directly lead to an increase in the FE runtime per frequency-configuration pair, currently standing at 3.3 sec. In contrast, such changes would affect primarily the upfront QEP cost of the TM simulations, while having only a mild effect on the per frequency-configuration cost, currently standing at 0.08 sec.  

\section{Summary}

\noindent In this work, we develop a semi-analytical tool for the simulation of 1D wave scattering generated by an architected layer sandwiched between two semi-infinite homogeneous domains. By design, the scattering layer is constructed as a union of bonded segments cut from various 1D periodic media. This allows us to describe the wave motion in each segment via superposition of the ``left'' and ``right'' (propagating or evanescent) Bloch eigenstates, computed for the mother periodic medium. Previous studies have shown that the quadratic eigenvalue problem, which furnishes the required Bloch eigenstates, has precisely two roots and provides for a complete solution in a given segment of a periodic medium -- irrespective of its length or boundary conditions. To compute the scattered field generated by an architected layer, the interplay between individual Bloch eigenstates (stemming from dissimilar periodic media) is synthesized via the propagator matrix approach. Through numerical simulations, the method is shown to be in excellent agreement with finite element simulations. The proposed methodology can be used to expedite an optimal design of 1D wave manipulation systems, whose effectiveness to achieve a target functionality (e.g. minimizing transmission) depends not only on the dispersion characteristics of the underlying periodic media but also on the specifics of an architected layer including, among other items, the makeups of individual segments, their lengths, and impedance contrasts at segment interfaces. For the optimized rainbow trap example that features $O(10^6)$ trial filter configurations, the proposed method is shown to achieve a $40\times$ computational speedup relative to finite element simulations. When compared to the conventional ``linear'' design of rainbow trap systems, the optimized layer -- obtained by combining the current approach with genetic algorithm search -- is found to yield a $40\%$ reduction in energy transmissibility over the target frequency band. The proposed scheme is equally applicable to an optimal design of 1D wave manipulation devices catering for other physical criteria (e.g. energy harvesting) or those governed by other wave-like equations, e.g. flexural wave (i.e. dynamic beam) equation. 

\bibliographystyle{plain}\bibliography{ref} 

\end{document}